\documentclass[12pt,a4paper]{article}
\usepackage{epsfig, latexsym, indentfirst}

\let\s=\sigma
\let\nn=\nonumber

\def\be{\begin{equation}}
\def\ee{\end{equation}}
\def\bea{\begin{eqnarray}}
\def\eea{\end{eqnarray}}
\def\ba{\begin{array}}
\def\ea{\end{array}}
\def\bi{\begin{itemize}}
\def\ei{\end{itemize}}

\author{
P. Petreczky\footnote{petr@cleopatra.elte.hu}\,,
Zs. Sz\'ep\footnote{szepzs@hercules.elte.hu} 
\,and 
A. Patk\'os\footnote{patkos@ludens.elte.hu} \\
\makebox[0.5cm]{}Department of Atomic Physics, E\"otv\"os University,\\
\qquad H-1117 P\'azm\'any P\'eter s\'et\'any 1/A, Budapest Hungary 
}
\title{
\huge \bf Coupled gap equations for the screening masses 
in the SU(2) Higgs model
}
\begin{document}
\maketitle
\begin{abstract}
\noindent
The complete set of static screening masses is determined
for the $SU(2)$ Higgs model from one-loop coupled gap equations. 
Results from the version, containing scalar fields both in the fundamental 
and adjoint representations are compared with the model arising when the 
integration over the adjoint scalar field is performed. 
A non-perturbative and non-linear mapping between the couplings of the
two models is proposed, which exhibits perfect decoupling of the heavy
adjoint scalar field. Also the alternative of a gauge invariant mass 
resummation is investigated in the high temperature phase.
\end{abstract}

\vskip 1truecm
\section{Introduction}
The finite temperature SU(2) Higgs model was extensively studied in
recent years in connection with the electroweak phase transition (EWPT)
and baryon asymmetry generation in the standard model (see Ref.
\cite{rew} for a review). Considerable progress was achieved 
in understanding the thermodynamics of the phase transition   
with the help of the method of dimensional reduction. In this approach the
superheavy modes (i.e. the non-zero Matsubara modes with typical
mass $\sim 2 \pi T$) and the heavy $A_0$ field (with a mass $\sim g T$)
are integrated out and the thermodynamics is described by an effective
theory, the 3d SU(2) Higgs model \cite{jako94,kaja94,kaja96}. The
properties of the phase transition and the screening masses were
studied in great detail using lattice Monte-Carlo simulations
of the reduced model 
\cite{kajanp,kajantielett,karsch96,karsch97,gurtler97} 
and also by 
Dyson-Schwinger (DS) technique in the full 4d theory
\cite{zwirner,fodor94} as well as in the effective 3d theory
\cite{bp94}.  
Lattice Monte-Carlo simulations predict that the line of first order 
transitions ends for some Higgs mass $m_H=m_H^c$
\cite{kajantielett,karsch97,gurtler97}. 
The same conclusion was obtained using the DS approach in
Ref. \cite{bp94} and the value of the critical mass
$m_H^c$ was found to be close to the prediction of Monte-Carlo
simulations.
Though, the validity of 
one-loop gap equations
was critically questioned \cite{jackiw96,cornwall98}, 
a recent two-loop calculation
\cite{eberlein98} has demonstrated that it is not an accident that 
the results of the one-loop level
analysis are fairly close to the conclusions of the numerical simulations.

The possibility of dimensional reduction is based on the fact that in the
full model  there are different well separated mass scales 
$g^2 T \ll g T \ll 2 \pi T$  for small couplings $g$. Recent 4d Monte-Carlo 
simulations of the finite temperature $SU(2)$ Higgs model 
\cite{fodor1,fodor2} provide good 
non-perturbative tests for the validity of dimensional reduction. 
A detailed discussion of relating 4d and 3d results
was published very recently in Ref. \cite{laine_new}.

The purpose of the present paper is twofold. First, we would like to provide 
some non-perturbative evidence for the decoupling of the $A_0$ field from the
gauge + higgs dynamics in the vicinity of the phase transition. 
We are going to solve a coupled set of gap equations for 
the 3d fundamental + adjoint Higgs model. This model emerges when the 
non-static modes are integrated out in the full finite temperature Higgs 
system. Its predictions for the screening masses will be compared 
with those obtained by Buchm\"uller and Philipsen (BP)
\cite{bp94} in the 3d Higgs model (with only one scalar field in the
fundamental representation) using the same technique. The main result of 
our investigation is a proposition for a non-perturbative and non-linear
mapping between the two models ensuring
quantitative agreement between the screening masses in a wide temperature
range on both sides of the transition. This high quality evidence for 
the decoupling of the $A_0$ field at the actual finite mass ratios presumes, 
however,
the knowledge of the "exact" value of the Debye screening mass, since for the 
proposed mapping its non-perturbatively determined value turns out to be 
essential.

Second, we wish to investigate the symmetric phase in more detail. 
There the Higgs and the Debye screening masses are both of the same order 
of  magnitude $\sim gT$ and thus in that regime there is no {\it a priori} 
reason for the $A_0$ field to 
decouple. This circumstance makes the quantitative relation of the screening 
masses calculated in the 3d fundamental + adjoint Higgs model particularly 
interesting in the high-T phase. Here we are going to apply two different 
resummation techniques and check to what extent 
persists a non-perturbative mass hierarchy in this part of the spectra.

All calculations of this paper are performed at 1-loop accuracy, but the 
above mentioned signal \cite{eberlein98}
for the good numerical convergence of the masses determined in the DS-scheme
gives us confidence that the effects we find will appear also in improved
treatments.

The presentation of our investigation proceeds as follows: 
in Section 2 we derive the coupled set
of gap equations for the 3d fundamental + adjoint Higgs model 
and discuss some problems related to the formal decoupling of
the adjoint Higgs-field, when its screening mass goes to  infinity.
In Section 3 we solve the coupled set of these equations numerically and
estimate the variation in the screening masses and some critical parameters
due to the presence of the adjoint Higgs field. In Section 4 we study the 
screening masses using an alternative gauge invariant resummation scheme, 
restricted in applicability to the symmetric phase. Finally, Section
5 presents our conclusions.

\section{The extended gap equations}
The Lagrangian of the three dimensional $SU(2)$ fundamental $+$ adjoint Higgs 
model is \cite{bp94,kaja94}
\bea\label{l3d}
L^{3D}&=&Tr\bigl[\frac{1}{2}F_{ij}F_{ij}+(D_i\Phi)^+(D_i\Phi)+\mu^2\Phi^+\Phi+
2\lambda (\Phi^+\Phi)^2\bigr]\nonumber\\
&&
+\frac{1}{2}(D_i\vec{A}_0)^2+
\frac{1}{2}\mu_D^2{\vec{A}_0}^2+{\lambda_A\over 4}(\vec{A}_0^2)^2+
2c\vec{A}_0^2Tr\Phi^+\Phi,
\eea
where
\bea
\Phi=\frac{1}{2}(\sigma 1+i\vec{\pi}\vec{\tau}), \quad 
D_i\Phi=(\partial_i-igW_i)\Phi,\quad W_i=\frac{1}{2}\vec{\tau}\vec{W}_i.
\eea
The relations between the parameters of the 3d theory  and those of the
4d theory are derived at 1-loop level perturbatively \cite{kaja94}:
\bea
&
g^2=g^2_{4d}T,\qquad\quad
\lambda=\left(\lambda_{4d}+\frac{3}{128\pi^2}g^4_{4d}\right)T,\qquad\quad
\lambda_A=\frac{17}{48\pi^2}g^4_{4d}T,\nonumber\\
&
c=\frac{1}{8}g^2_{4d}T,\qquad\quad
\mu_D^2=\frac{5}{6}g^2_{4d}T^2,\qquad\quad
\mu^2=\left(\frac{3}{16}g^2_{4d}+\frac{1}{2}\lambda_{4d}\right)T^2-
\frac{1}{2}\mu^2_{4d}.
\eea 
If the integration over the $A_0$ adjoint Higgs field is performed we obtain
the model investigated in \cite{bp94} with parameters $\bar g, \bar\lambda,
\bar\mu$. These couplings of the reduced theory are related to the
parameters of the 3d fundamental $+$ adjoint Higgs theory through the
following relations:
\be
\label{eq:relpar1}
\bar g^2=g^2\left(1-\frac{g^2}{24\pi \mu_D}\right),\qquad
\bar \lambda=\lambda-\frac{3c^2}{2\pi\mu_D},\qquad
\bar\mu^2=\mu^2-\frac{3c\mu_D}{2\pi}.
\label{pertmap}
\ee   
In particular, we note that the $\bar\mu$ scale serves as the temperature 
scale of the fully reduced system, while $\mu$ is the scale for the
system containing both the fundamental and the adjoint scalars. The two are 
related perturbatively by a constant shift.

In order to perform the actual calculations in the broken phase it is 
necessary to shift 
the Higgs field, $\sigma \rightarrow v + \sigma'$. After this shift  
and the gauge-fixing (the gauge fixing parameter is denoted by
$\xi$) the Lagrangian including the ghost terms assumes the form
\bea\label{lexplit}
L &=& {1\over 4}\vec{F}_{\mu\nu}\vec{F}_{\mu\nu} + {1\over 2\xi}
(\partial_{\mu}\vec{W}_{\mu})^2 + {1\over 2}m_0^2\vec{W}_{\mu}^2 \nn\\
&& + {1\over 2}(\partial_{\mu}\s')^2 + {1\over 2} M_0^2 \s'^2
   +{1\over 2}(\partial_{\mu}\vec{\pi})^2 + \xi{1\over 2}m_0^2\vec{\pi}^2\nn\\
&& +{g^2\over 4}v\s'\vec{W}_{\mu}^2 + {g\over 2}\vec{W}_{\mu}\cdot(
    \vec{\pi}\partial_{\mu}\s'-\s'\partial_{\mu}\vec{\pi})
   + {g\over 2}(\vec{W}_{\mu}\times\vec{\pi})\cdot\partial_{\mu}\vec{\pi}\nn\\
&& + {g^2\over 8}\vec{W}_{\mu}^2(\s'^2+\vec{\pi}^2)
   + \lambda v\s'(\s'^2+\vec{\pi}^2)+{\lambda\over 4}(\s'^2+\vec{\pi}^2)^2\nn\\
&& + {1\over 2 } {(D_i \vec{A_0})}^2+{1\over 2} m_{D0}^2 {\vec{A_0}}^2
   + {\lambda_A\over 4} {({\vec{A_0}}^2)}^2+2 c v \s' {\vec{A_0}}^2
   + c {\vec{A_0}}^2 ({\s'}^2+{\vec{\pi}}^2)\nn\\
&& + \partial_{\mu}\vec{c^*}\partial_{\mu}\vec{c}
   + \xi m_0^2 \vec{c^*}\vec{c}\nn\\
&& + g\partial_{\mu}\vec{c^*}\cdot(\vec{W}_{\mu}\times\vec{c})
   + \xi{g^2\over 4}v\s'\vec{c^*}\vec{c}
   + \xi{g^2\over 4}v\vec{c^*}\cdot(\vec{\pi}\times\vec{c})
   + {1\over 2}\mu^2 v^2 + {1\over 4}\lambda v^4\nn\\
&& + {1\over 2}(\mu^2+\lambda v^2)(\s'^2+\vec{\pi}^2)
   + v(\mu^2 +\lambda v^2)\s'\ ,
\label{shiftL}
\eea
where the following notations were introduced for the tree-level masses: 
$m_0^2={1\over4} g^2 v^2$ (the vector boson mass), $M_0^2=\mu^2+3 
\lambda v^2$ (the Higgs mass) and $m_{D0}^2=\mu_D^2+2 c v^2$ (the Debye mass). 
The last two terms of (\ref{shiftL}) 
arise from the Higgs potential after the shift in the Higgs 
field $\s$. For $\mu^2 < 0$, they vanish if one expands around the classical 
minimum $v^2 = -\mu^2/\lambda$. In general, however, these terms have to be 
kept \cite{bp94}.

In order to obtain the coupled gap equations one replaces the tree-level masses
by the exact masses
\be
m_0^2\rightarrow m^2+\delta m^2,~~~~~ M_0^2 \rightarrow
M^2+\delta M^2,~~~~~m_{D0}^2\rightarrow m_D^2+\delta m_D^2,
\ee
and treats the differences 
$\delta m^2=m_0^2-m^2,~~\delta M^2=M_0^2-M^2,~~\delta m_D^2=m_{D0}^2-m_D^2$
as counterterms. The exact Goldstone and ghost masses are both equal to
$\sqrt{\xi} m$, where $m$ is the exact gauge boson mass.
The gauge invariance of the self-energies of the Higgs and gauge
bosons is ensured by introducing appropriate vertex resummations. 
Their explicit formulae can be found in \cite{bp94}. In the present 
extended model, a resummation of the Higgs-$A_0$
vertex would be also necessary if the gauge invariance of the $A_0$
self-energy is to be ensured. Then the only source of the gauge 
dependence which would remain is the equation for the vacuum expectation 
value $v$.

All these resummations are equivalent to work with 
the following gauge invariant Lagrangian: 
\bea
L^{3D}_I & = & \frac{1}{4}\vec{F}_{ij}\vec{F}_{ij}+
Tr\left((D_i\Phi)^+D_i\Phi-\frac{1}{2}M^2\Phi^+\Phi\right)+
\frac{1}{2}(m_D^2-\frac{8cm^2}{g^2})\vec{A}_0^2+\nonumber\\
&&
\frac{g^2M^2}{4m^2}Tr(\Phi^+\Phi)^2+
2c\vec{A}_0^2Tr\Phi^+\Phi.
\eea
In this Lagrangian one shifts the Higgs field around its classical 
minimum $\s\rightarrow \s' + {2 m \over g}$ and adds the corresponding gauge 
fixing and ghost terms \cite{bp94}. Shortly, we shall argue that the 
$A_0$-Higgs vertex resummation arising from the replacement of $v$ by 
$2m/g$ when the scalar field is shifted in the last term of the above 
Lagrangian destroys the mass-hiearchy between the heavy 
$A_0$ and the light gauge and Higgs fields. Therefore in this paper we have to 
give up the full gauge independence of the resummation scheme. 
The numerical solution to be presented below shows that the gauge dependence
of the $A_0-\Phi$ vertex
in our resummation scheme introduces only a minor additional gauge dependence 
beyond that of the equation for the
vacuum expectation value \cite{bp94} appearing below in Eq. (\ref{eq:gapeq4}). 

The coupled set of gap equations is constructed from that of Ref.
\cite{bp94} by adding the contributions due to the presence of the 
adjoint Higgs field. The self-energy contributions for the 3d fundamental 
Higgs and for the 3d adjoint Higgs model were already calculated  in 
\cite{bp94} and \cite{gap}, respectively. 
Below we list only the additional contributions to the 
self-energies, which all contain at least one $A_0-\Phi$ vertex (the
corresponding diagrams are listed in the Appendix A). We emphasize
once again that no resummation of the $A_0-\Phi$ vertex was applied.

The additional contribution to the self-energy of the $A_0$ 
field coming from Higgs, Goldstone, gauge and ghost fields (diagrams {\bf a-i}) 
is
\bea
\delta\Pi_{A_0}^{H,G,gh}(p,m,M,m_D) & = & -{4 c v^2(\mu^2+\lambda v^2)\over M^2}+
{3 c g v\over \pi} \left( {M\over 4 m}+{m^2\over M^2} \right)-{c M
\over 2 \pi}
+{3 \sqrt{\xi} \over 4 \pi} (g v -2 m)\nonumber\\
&&
+{4 c^2 v^2\over \pi}\left[
{3\over 2} {m_D\over M^2}-
\frac{1}{p}\arctan\frac{p}{m_D+M}\right].
\eea
There is also an additional contribution  
to the gauge boson self-energy  coming from the adjoint Higgs field 
(diagram {\bf m}):
\be
\delta\Pi_T^H(p,m,M,m_D)=\frac{3 c g}{2 \pi}\frac{m v}{M^2}m_D.
\ee
The contribution of $\vec{A}_0$ to the Higgs self-energy
(diagrams {\bf j-l}) is the following
\bea
\delta\Pi_H^{A_0}(p,m,m_D)=-\frac{3 m_D
c}{2 \pi}-\frac{6 c^2 v^2 }{\pi}
\frac{1}{p}\arctan\frac{p}{2m_D}+{9 gcv \over 4\pi m} m_D.
\eea

Making use also of the pieces of the self-energies calculated in 
\cite{bp94,gap} we write
down a set of coupled on-shell gap equations for the screening masses of the
magnetic gauge bosons, fundamental Higgs and adjoint $A_0$ fields in the 
form
\bea
\label{eq:gapeq2}
m^2&=&\Pi_T(p=im,m,M)+\delta\Pi_T^{A_0}(p=im,m_D)+\delta\Pi_T^{H}
(p=im,m,M,m_D),\\
\label{eq:gapeq3}
M^2&=&\Sigma(p=iM,m,M)+\delta\Pi_H^{A_0}(p=iM,m,m_D),\\
\label{eq:gapeq1}
m_D^2&=&\Pi_{00}(p=im_D,m,m_D)+\delta\Pi_{A_0}^{H,G,gh}(p=im_D,m,M,m_D),
\eea
where $\Pi_T$ and $\Sigma$ are defined by  eqs. (17),~(18) of Ref. 
\cite{bp94}. $\delta \Pi_T^{A_0}$ and $\Pi_{00}$ were presented in eqs. 
(7),(8) of \cite{gap}.

If on the right hand side of the third equation one inserts the tree
level masses, the next-to-leading order result of Ref.\cite{rebhan} is recovered
for the Debye mass in the $SU(2)$ Higgs model. 
 
The equation for the vacuum expectation value makes the set of the above 
three equations complete:
\be
v(\mu^2+\lambda v^2)=\frac{3}{16\pi}g\left(4m^2+\sqrt{\xi} M^2+\frac{M^3}{m}
\right)+\frac{3c}{2 \pi}v m_D.
\label{eq:gapeq4}
\ee
It is important to notice that this equation can be rewritten as
\be\label{vac2}
v(\mu_{eff}^2+\lambda v^2)=\frac{3}{16\pi}g\left(4m^2+\sqrt{\xi} 
M^2+\frac{M^3}{m}\right),
\ee
with 
\be
\mu^2_{eff}=\mu^2-\frac{3c}{2 \pi}m_D.
\label{muscale}
\ee 
This equation is formally identical to the equation of BP for the vacuum
expectation value \cite{bp94}. On the basis of this observation, 
we expect that the main effect of the $A_0$ integration is the above shift 
in the $\mu^2$-scale. Since $m_D$ is itself a non-trivial function of $\mu$
this non-perturbative mapping is also nonlinear.

A very similar set of equations could be derived for the
case of the gauge invariant resummation of the $A_0-\Phi$ vertex.
They are listed in Appendix B. 
For instance, one would write in the last term on the right hand side
of (\ref{eq:gapeq4}) $2m/g$ on the place of $v$, which would
suggest a different 
redefinition of the temperature
($\mu^2$) scale:
\be
\tilde \mu^2_{eff}=\mu^2-{3 c\over \pi g}{m m_D\over v}.
\label{nonpertmap11}
\ee
If the tree level masses are inserted into this redefinition it gives
the usual relation between the mass parameters of the full static and
the $A_0$-reduced models (see Eq. (\ref{pertmap})).

\section{Numerical results}
The main goal of the present investigation is to propose a scheme of
solution for the full static Higgs model (\ref{l3d}) which reproduces the BP solution
of the reduced static model (with $A_0$ integrated out). The 
existence of such a solution is made plausible by the Appelquist-Carazzone
theorem \cite{apcar}, but by no means it is trivial to construct it for two
obvious reasons. The decoupling theorem is valid only for infinitely
different mass scales, while the $m_D/m$, $m_D/M$ ratios are finite in the
realistic case. There are corrections to the theorem even if we would be
able to compare the exact values of the corresponding masses calculated in
the two models for the perturbatively related values of the couplings. 
The second source of deviations comes from the resummation applied in the
process of the perturbative solutions. It is not clear which resummed solution 
of the full static model would correspond to the BP-resummed approximate
solution of the reduced 3d effective model at one-loop level. 

Though the construction of a good quality correspondence is a very difficult 
task, it is a necessary effort if one wishes to go beyond the "existence proof" 
of the decoupling in case of the resummed solutions.

We have to admit that it would be much easier to assess the status of
$A_0$-decoupling and the quality of the BP-solution if the exact (Monte-Carlo) 
solution of the model (\ref{l3d}) would be available.  
However, Monte-Carlo simulations of the gauge + fundamental + adjoint Higgs
system are extremely difficult to realize (see discussion in Ref.
\cite{kajanp}). 
Therefore our present
construction can be considered a first detailed attempt to establish
quantitative arguments for the $A_0$-decoupling.

\begin{figure}[t]
\epsfysize=7cm
\epsfxsize=10cm
\centerline{\epsffile{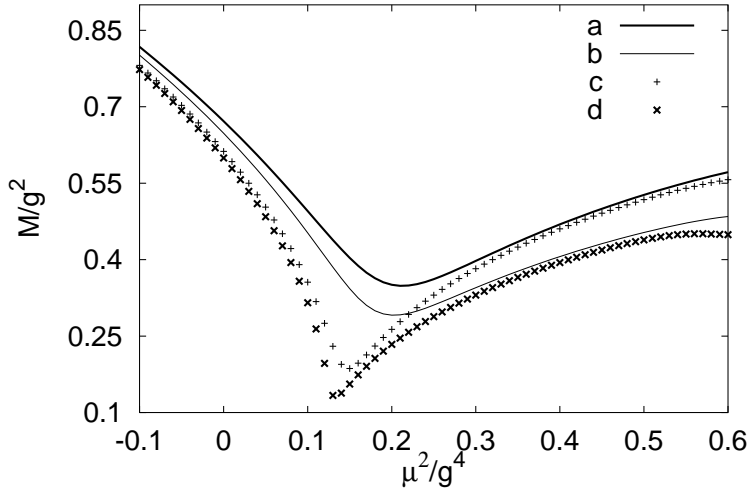}}
\caption{\small
The Higgs mass in units of $g^2$ as function of $\mu^2/g^4$ calculated at
$\lambda/g^2=1/8$ using the gauge-invariant
$A_0-\Phi$ vertex resummation version of the gap equations. Shown are the
Higgs mass derived in the full static theory in the Landau- (a) and in the
Feynman-gauge (b) and the Higgs mass values in the $A_0$-reduced theory in
the Landau-gauge (c) and in the Feynman-gauge (d). The $\mu^2$-shift indicated
 by  eq.(\ref{pertmap}) was applied.}
\end{figure}

Our first attempt at solving the full static model followed the gauge
invariant vertex resummation procedure employed also by the BP solution of the 
$A_0$-reduced model. In Fig.1 the results of the two solutions for the Higgs
mass $M$ are displayed taking into account the perturbative mapping
(\ref{pertmap}) between the parameters of the two models. The deviations
are large, especially in  the critical region. We arrived at a negative
conclusion:
The gauge invariantly resummed one-loop solutions of the gap equations of the
two models do not correspond to each other if the perturbative
$A_0$-integration is correct. 

We have also tried to compare the 
predictions of the full
static and the reduced models in the case when the mass parameter of
the reduced model is chosen according Eq. (\ref{nonpertmap11}).
Such non-perturbative mapping between the parameters of the two models
improves somewhat the situation deep in the broken phase, however,
near the crossover region the values of the masses calculated in the
two models differs considerably. We conclude that if gauge invariant
resummation of the  $A_0-\Phi$ vertex is used we are not able to
find a physically motivated relation between the parameters of the full
static and the reduced models
with the help of which the two 
models give acceptably close mass predictions. Therefore we will not
discuss further the fully gauge invariant resummation scheme but turn 
to the discussion of the results obtained in the case when the 
$A_0-\Phi$ vertex left unresummed.

If the $A_0-\Phi$ vertex left unresummed, a very simple expectation
emerges concerning the effect of the $A_0$ integration on the mass spectra,
as it was disscused on the basis of (\ref{muscale}) in the previous section.  
Therefore,
we will first compare the predictions for the Higgs and gauge boson masses
from the coupled gap equations  (\ref{eq:gapeq2})-(\ref{eq:gapeq4}) 
of the 3d fundamental + adjoint Higgs model with
those obtained in the $A_0$-reduced theory, the 3d Higgs model \cite{bp94}. The
corresponding Higgs masses are shown in Fig. 2 using two different gauges. 
The results obtained in the $A_0$-reduced theory are displayed after the shift
required by eq.(\ref{pertmap}) is performed. As
one can see the difference between the full and the reduced theory is still
visible in the vicinity of the crossover. In this region the relative 
difference between the predictions of the full and the reduced theory
is about $20 \%$. 

\begin{figure}
\epsfysize=7cm
\epsfxsize=10cm
\centerline{\epsffile{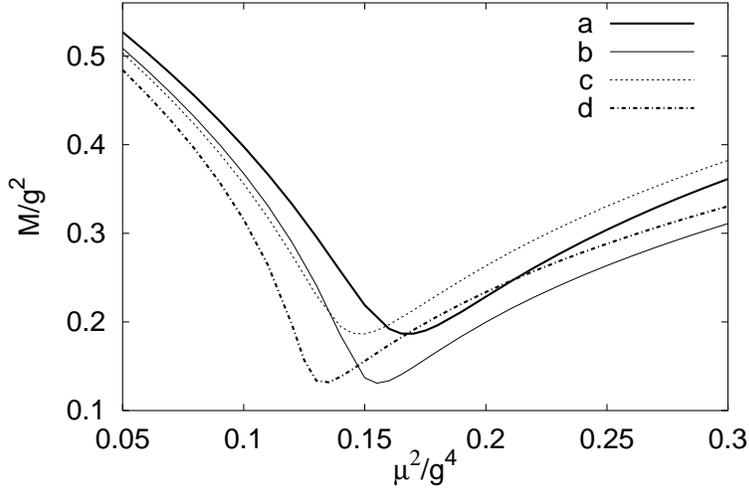}}
\caption{\small The Higgs boson masses at $\lambda/g^2=1/8$ (crossover region) 
in units of $g^2$ as function of $\mu^2/g^4$ in the 3d fundamental + adjoint 
Higgs model and in the 3d SU(2) Higgs ($A_0$-reduced) theory. Shown are the 
Higgs mass in the full static theory in the $\xi=0$ (Landau) gauge (a) and in 
the $\xi=1$ (Feynman) gauge (b), and the Higgs boson mass in the $A_0$-reduced
theory in the $\xi=0$ gauge (c) and in the the $\xi=1$ gauge (d).}
\end{figure}
Our proposal to resolve this relatively 
large deviation is to introduce a more complicated relationship between
the couplings. Having gained intuition from eq.(\ref{muscale}), 
we have plotted the mass-predictions for the Higgs-field
derived from our full set of equations against the results of BP calculated
for couplings taken from (\ref{pertmap}) with a replacement $\mu_D\rightarrow
m_D$:
\be 
g_{eff}^2=g^2(1-{g^2\over 24\pi m_D}),\qquad
\lambda_{eff}=\lambda-{3c^2\over 2\pi m_D},\qquad 
\mu_{eff}^2=\mu^2-{3c\over 2\pi}m_D.
\label{nonpertmap}
\ee  
The non-trivial nature of this replacement becomes clear from Fig.3
where the $\mu^2$-dependence of $m_D$ is displayed. Clearly, its
non-trivial $\mu^2$-dependence is most expressed in the neighbourhood 
of the phase transformation (crossover) point $\mu^2/g^4 \in (0.1-0.2)$. 
The application of this
mapping to the data obtained from the model containing both the fundamental 
and the adjoint representation leads to a perfect agreement of the two data 
sets for large values of $\lambda /g^2$. For smaller values of $\lambda /g^2$ 
(1/32,1/64) the mapping (\ref{nonpertmap}) works very well in the symmetric 
phase, but in the broken phase (\ref{pertmap}) seems to be the better choice.

We suspect, that the tree level piece in $m_D$ arising from the Higgs-effect,
should not be included into the correction of (\ref{pertmap}), since it is 
itself a tree-level effect. Therefore we propose the following replacement in
(\ref{nonpertmap}):
\be
m_D\rightarrow \sqrt{m_D^2-2cv^2}.
\label{nonpertmap_new}
\ee
In Fig.4 it is obvious that a very good agreement could be obtained with this
mapping between the Higgs mass predictions of the one-loop
gap equations of the full static and the $A_0$-reduced theory for $\lambda /g^2
=1/32$. The quality of the agreement on both sides of the phase transition
is good, signalling that the influence of the ``mini-Higgs'' effect in the 
symmetric phase is negligible. Therefore it is not surprising that for
$\lambda /g^2=1/8$ the same quality of agreement is obtained like before.

It is important to notice that there is a strong
gauge parameter dependence in the symmetric phase and in the vicinity
of the crossover. The variations due to the change in the gauge 
are equal  in the full and in the 
reduced theory, which indicates that the additional gauge dependence,
introduced by the gauge non-invariant resummation of the $A_0$ field is
negligible. The mapping (\ref{nonpertmap_new}) performs equally well
in Landau- and Feynman-gauges.

Other quantities which are worth of considering for the comparison 
of the full 3d and the reduced theories are $\lambda_c/g^2$, the endpoint of 
the first order transition line and $\mu_{+}/g^2$, the mass parameter above 
which the broken phase is no longer metastable. 
The  values of $\mu_{+}^2/g^4$ for different scalar couplings and different
gauges in the full and in the reduced theory are summarized in Table 1.
Here the mapping (\ref{nonpertmap}) could be implemented only
by extrapolating from smaller $\mu^2/g^4$, since the
end-points of metastability do not correspond to each other, and in some
cases $m_D$ could not be determined from the gap equations. Also here
for larger values of $\lambda/g^2$ the application of (\ref{nonpertmap}) led
to an improved agreement between the end-point $\mu_+^2/g^4$ values, while for 
$\lambda /g^2 =1/48,1/64$ the mapping (\ref{pertmap}) works better. 
In the table
we have displayed $\mu_+^2/g^4$ values of the $A_0$-reduced theory shifted 
perturbatively and with help of the best performing non-perturbative
mapping (\ref{nonpertmap_new}). For both gauges the latter agrees with the
$\mu_+^2/g^4$-values of the full static theory very well.
\begin{center}
\begin{tabular}{|l|l|l|l|}
\hline
$~~~~\lambda/g^2~~~$  &$~~~~~~~~~~~~~A~~~~~~~~~~$
& $~~~~~~~~~~~~~B~~~~~~~~~~$ & $~~~~~~~~~~~~~C~~~~~~~~~~$\\
\cline{2-4}
$~~~~~~~~~~~~~~$ & $~~~\xi=0~~~$ \vline $~~~\xi=1~~~$
& $~~~\xi=0~~~$ \vline $~~~\xi=1~~~$ 
& $~~~\xi=0~~~~$\vline $~~~~\xi=1~~~$
\\
\hline
$~~~1/32~~~$  &$~~~0.1516~~~~~~~0.1423~~~$
&$~~~0.1426~~~~~~0.1341~~~$
&$~~~0.1499~~~~~~~0.1405~~~$\\
$~~~1/48~~~$  &$~~~0.1647~~~~~~~0.1558~~~$
&$~~~0.1627~~~~~~0.1541~~~$
&$~~~0.1637~~~~~~~0.1546~~~$\\
$~~~1/64~~~$  &$~~~0.1841~~~~~~~0.1750~~~$
&$~~~0.1881~~~~~~0.1808~~~$
&$~~~0.1875~~~~~~~0.1792~~~$\\
\hline
\end{tabular}
\vskip0.3truecm
Table.~1: {{\small Values of $\mu_{+}^{2}/g^4$ in the full static theory (A), 
in the perturbatively reduced theory (B) and in the reduced theory
obtained using non-perturbative matching described in the text (C). 
Calculations were done 
in the Landau ($\xi =0$) and in the Feynman ($\xi=1$) gauges.}}
\end{center}

\begin{figure}
\epsfxsize=9cm
\epsfysize=6cm
\centerline{\epsffile{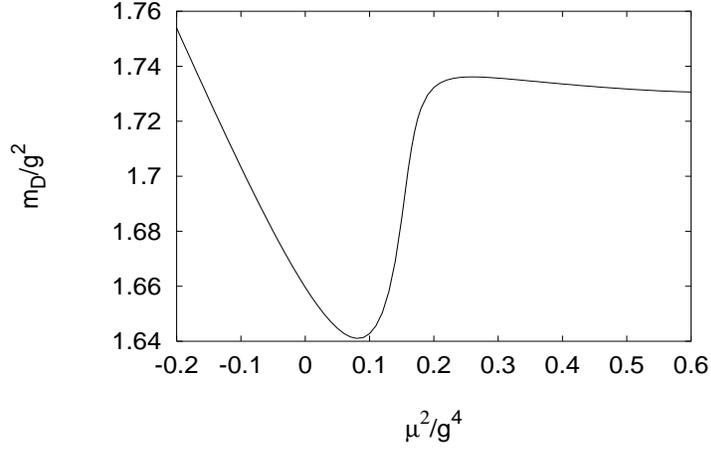}}
\caption{\small
The $\mu^2$ dependence of the Debye mass for
$\frac{\lambda}{g^2} =1/8$.}
\end{figure}

The endpoint of the $1^{st}$ order line in the Landau gauge in the 
3d Higgs theory was found at $\lambda_c/g^2=0.058$. The
corresponding critical scalar coupling in the full 3d theory is within
the 1\% range. In Feynman gauge we find $\lambda_c/g^2=0.078$ for the
$A_0$-reduced theory and the
corresponding value for the full 3d theory lies again very close to it. 
Thus the $A_0$ field has almost no effect on the position of the endpoint.
The strong gauge dependence of $\lambda_c$ indicates, however, that
higher order corrections to this quantity are important.

\begin{figure}
\epsfxsize=10cm
\epsfysize=7cm
\centerline{\epsffile{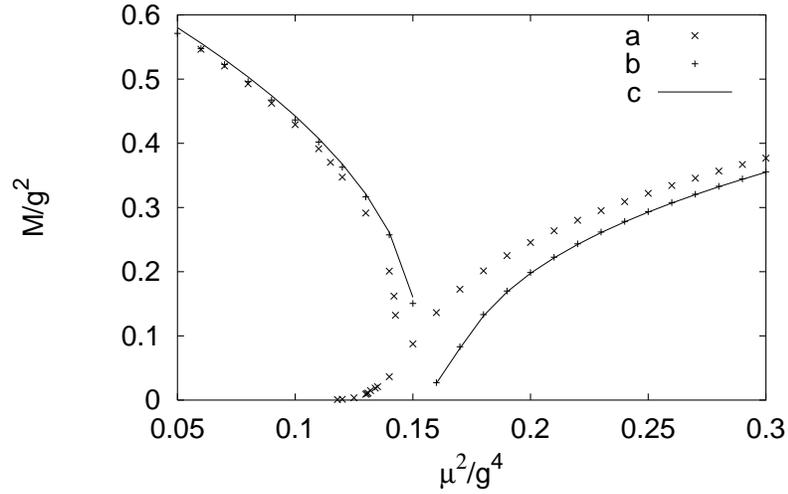}}
\caption{\small
The Higgs boson mass in units of $g^2$ as function of $\mu^2/g^4$
calculated at $\lambda/g^2=1/32$ in the Landau gauge
in the full static theory and in the
$A_0$-reduced theory. Shown are the Higgs mass in the reduced theory
obtained
by perturbative reduction (a), in the reduced theory obtained by
non-perurbative
matching (cf. eqs. (\ref{nonpertmap}), (\ref{nonpertmap_new})) (b)  and
in full static theory (c).}
\end{figure}

The gauge dependence of the screening masses is even more pronounced deep 
in the symmetric phase
($\mu^2/g^4>0.3$). For example the value of the gauge boson mass is roughly
$0.28 g^2$ in the symmetric phase for the Landau gauge. The
corresponding value in the Feynman gauge is about $0.22 g^2$. The gauge
dependence of the gauge boson mass is somewhat weaker at the 2-loop
level \cite{eberlein98}. It should be also noticed that the gauge boson
mass depends weakly on the parameters of the scalar sector
($\mu,~\mu_D,~\lambda,~\lambda_A$). This fact was also noticed in
previous investigations \cite{bp94,gap}.

\section{Screening masses in the symmetric phase with 
a gauge invariant resummation scheme}

The main motivation for the present investigation was to gain  insight
into the decoupling of the dynamics of the fundamental and 
the adjoint Higgs fields. The degree of the decoupling is 
expected to depend on the mass ratio of the fundamental
and adjoint Higgs fields. In the symmetric phase both
masses are of the same order in magnitude (eg. $\sim gT$). 
Therefore the hierachy of the $A_0$ and Higgs masses 
can only be present  due to numerical prefactors. The persistence 
of the perturbatively calculated ratio should be checked in any 
non-perturbative approach.

As we have seen in the previous section the gauge dependence in the
symmetric phase is too strong in the applied schemes to give a stable 
estimate for the mass ratio of the fundamental and the adjoint Higgs fields.
A reliable non-perturbative estimate for the Higgs mass deep in the symmetric
phase (defined through the
pole of the propagator) is even more interesting because it was not
measured so far on lattice. Therefore, in this section
we will investigate a coupled set of gap equations in the symmetric phase
which is based on the gauge invariant resummation scheme of 
Alexanian and Nair (AN) \cite{alex}. In this approach one can avoid any vacuum 
expectation value for the Higgs field in
the symmetric phase and because of this fact this approach is gauge
invariant.

In order to derive the 1-loop gap equations for the Higgs model in the
AN scheme one has to add the following terms to the original Lagrangian:
\be
\delta L  = {1\over 2} m^2 A_i (\delta_{ij}+{\partial_i \partial_j\over
\partial^2}) A_j+ f^{abc} V_{ijk} A_i^a A_j^b A_k^c
-{1\over 2 \xi} \partial_i A_i (1-m^2{1\over \partial^2}) \partial_j
A_j.
\label{dlan}
\ee
The first term in this expression is the mass term, the second
corresponds to a specific vertex resummation, where the explicit expression 
for $V_{ijk}$ could be find in Ref. \cite{alex}. Finally, the last term
is the gauge fixing term. For the coupled gap equations one has to reevaluate 
those self-energy diagrams of the gauge, Higgs and $A_0$ fields which 
involve the modified gauge propagators from (\ref{dlan}). 
Straightforward calculations lead to the following equations:
\begin{figure}
\epsfxsize=10cm
\epsfysize=7cm
\centerline{\epsffile{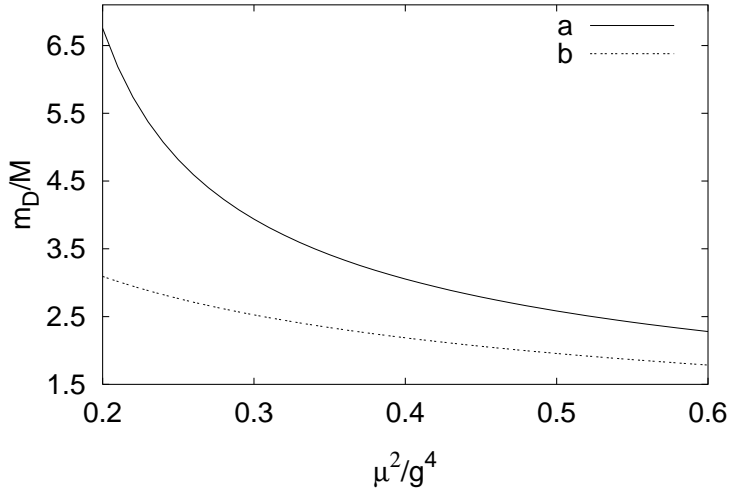}}
\caption{\small
The ratio of the Debye and the fundamental Higgs masses for
${\lambda\over g^2}=1/8$ calculated from gap eqs. (\ref{aneq}) (a) and
the leading order result (b).}
\end{figure}

\bea
&&
m^2=C g^2 m + {g^2 m\over 4 \pi} \biggl(2 f(m_D/m)+ f(M/m)\biggr),\nonumber\\
&&
M^2=\mu^2+{1\over 4 \pi} \biggl({3\over 4} g^2 M F(M/m)-6 \lambda M-6 c m_D
\biggr),\nonumber\\
&&
m_D^2=\mu_D^2+{1\over 4 \pi} \biggl( 2 g^2 m_D F(m_D/m)-5 \lambda_A m_D-
8 c M\biggr),
\label{aneq}
\eea
where $C={1\over 4\pi} (21/4 \ln3-1)$ \cite{alex} and the
following function were introduced
\bea
&&
f(z)=-{1\over 2} z+\bigl(z^2-{1\over 4}\bigr) {\rm arctanh}{1\over 2 z},\\
&&
F(z)=-1-{1\over z}+\bigl( 4 z - {1\over z} \bigr) \ln(1+ 2 z).
\eea

Let us first discuss the ratio of the $A_0$ and the fundamental Higgs
masses. In Fig.5 this ratio is shown as calculated from the eqs.
(\ref{aneq}) and compared with the corresponding perturbative value.
The $\mu$ interval in this plot corresponds to the temperature range 
relavant for the 
electroweak theory $T<1 TeV$.
We have also analyzed the $\mu$-dependence of the fundamental
Higgs mass alone in the full static and in the $A_0$-reduced model.
For $\mu^2/g^4$ in the interval $(0.2-0.3)$ the result of the gauge invariant
approach agrees fairly well with the masses obtained in the BP-scheme.
In Fig.6 the difference between the Higgs masses calculated 
from the coupled set of gap equations (\ref{aneq}) and the leading
order perturbative result ($M_0$) is shown. As one can see, 
the non-perturbative correction to the Higgs mass is the largest for
small $\mu$ and is decreasing as $\mu$ increases reaching the percent level
for large enough $\mu$.

\begin{figure}
\epsfxsize=10cm
\epsfysize=7cm
\centerline{\epsffile{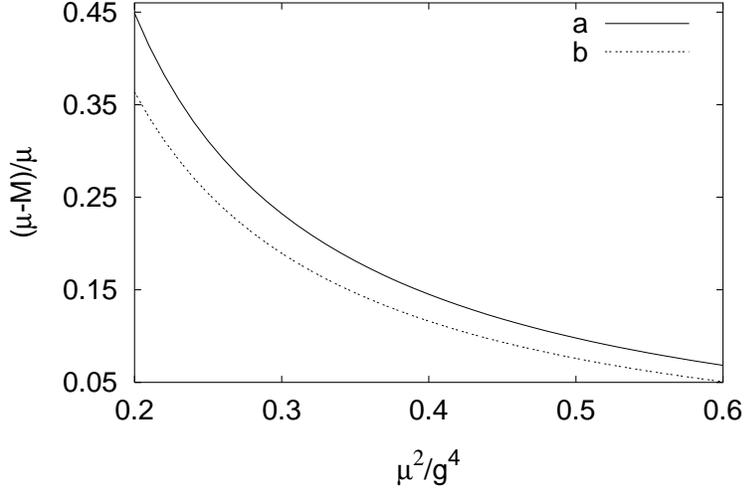}}
\caption{\small
The non-perturbative correction to the Higgs mass as
the function of $\mu$ calculated from the full static (a)
and from the $A_0$-reduced theory (b).}
\end{figure}

The relative difference between the full static and the $A_0$-reduced theory,
however, is slowly increasing as $\mu$ increases and the hierarchy between 
the $A_0$ and the Higgs masses becomes less pronounced as $\mu$
gets larger (see Fig.6).
The relative difference between the Higgs masses calculated in the 
full static and in the $A_0$-reduced theory varies between $20\%$ for 
$\mu^2/g^4=0.2$ and $35\%$ for
$\mu^2/g^4=0.6$.

It is also important to notice that the $A_0$ field is 
not sensitive to the dynamics of the Higgs field. In particular it
turns out that $m_D$ depends weakly on $\mu$ and $\lambda$ in the
symmetric phase and its value is close to the corresponding value 
calculated in 3d adjoint Higgs model. Let us notice that the magnetic
mass in this resummation scheme also seems to be insensitive to the
dynamics of scalars, therefore the magnetic and electric screening masses
are close to their values determined in the pure $SU(2)$ gauge model 
\cite{gap}.

\section{Conclusions}

The Appelquist-Carazzone (AC) theorem provides an important asymptotic
basis for the derivation of reduced effective models, when fields with
largely different masses appear in a field theoretical model. It states that
in the infinite mass limit the n-point functions of the light degrees of 
freedom can be calculated from an effective theory, in which the effect of the
heavy fields is present only in the couplings. In the electroweak theory
these effective models were determined perturbatively. In resummed
perturbation theory for finite orders the fulfillment of the theorem cannot
be checked on a diagram by diagram basis.

The comparative investigation of the screening masses of the full static
and the $A_0$-reduced theories of the finite temperature $SU(2)$ Higgs model
gives us a very valuable opportunity to study how well the AC theorem
works under realistic mass ratios. In particular in the symmetric phase
of the theory we have seen that a non-perturbative coupling relation 
(\ref{nonpertmap}) is necessary to map almost perfectly the masses determined
in the $A_0$-reduced model onto those found from the gap equations of the  
complete static effective model. The $\lambda/g^2$ -range (1/64-1/8) has covered
the regime of strong first order transitions to values where only
smooth crossover takes place. The correspondence between specific solution
schemes, which is compatible with the AC theorem represents constructive
evidence for the validity of the theorem.

The quality of the mapping did not depend on
the gauge-choice, which however, strongly influences the actual values of the
screening masses. Therefore, we have applied also a gauge-invariant
resummation scheme in the symmetric phase. The results show larger $m_D/M$
ratio than perturbatively predicted, which makes the basis for the $A_0$
reduction more solid.

In the broken symmetry phase the non-perturbative mapping as given by
(\ref{nonpertmap}) does not work. The attempt to separate the non-perturbative
change of the Debye-mass from the result of the symmetry breaking led us
to propose the mapping (\ref{nonpertmap_new}). It gave very satisfactory
results for both the Higgs mass and the upper metastability
edge $\mu_+/g^2$ in the Higgs-mass range $\lambda /g^2\in (1/32,1/8)$, when
resummed one-loop solutions of different relevant models in specific schemes
are calculated. We believe that our phenomenological observation opens the 
path towards a more refined physical understanding of the relationship of the 
couplings in the two models. 
This is necessary for the consolidation of the status of a non-perturbative 
$A_0$-decoupling from the static sector of the finite temperature 
Higgs theory.

\section*{Ackowledgements}The authors acknowledge informative discussion
on the subject of the paper with A. Jakov\'ac and O. Philipsen. We thank the 
Hungarian Research Fund (OTKA) for support.

\newpage

\section*{Appendix A} 
\vskip0.8truecm
Below we list graphycally the additional diagrams contributing to 
the $A_0$ ({\bf a-i}), Higgs boson ({\bf j-l}) and the vector
boson ({\bf m}) self-energies and the vacuum expectation value ({\bf n}).
\vskip1truecm
\unitlength1cm
\epsfysize=9cm
\centerline{
\epsfbox{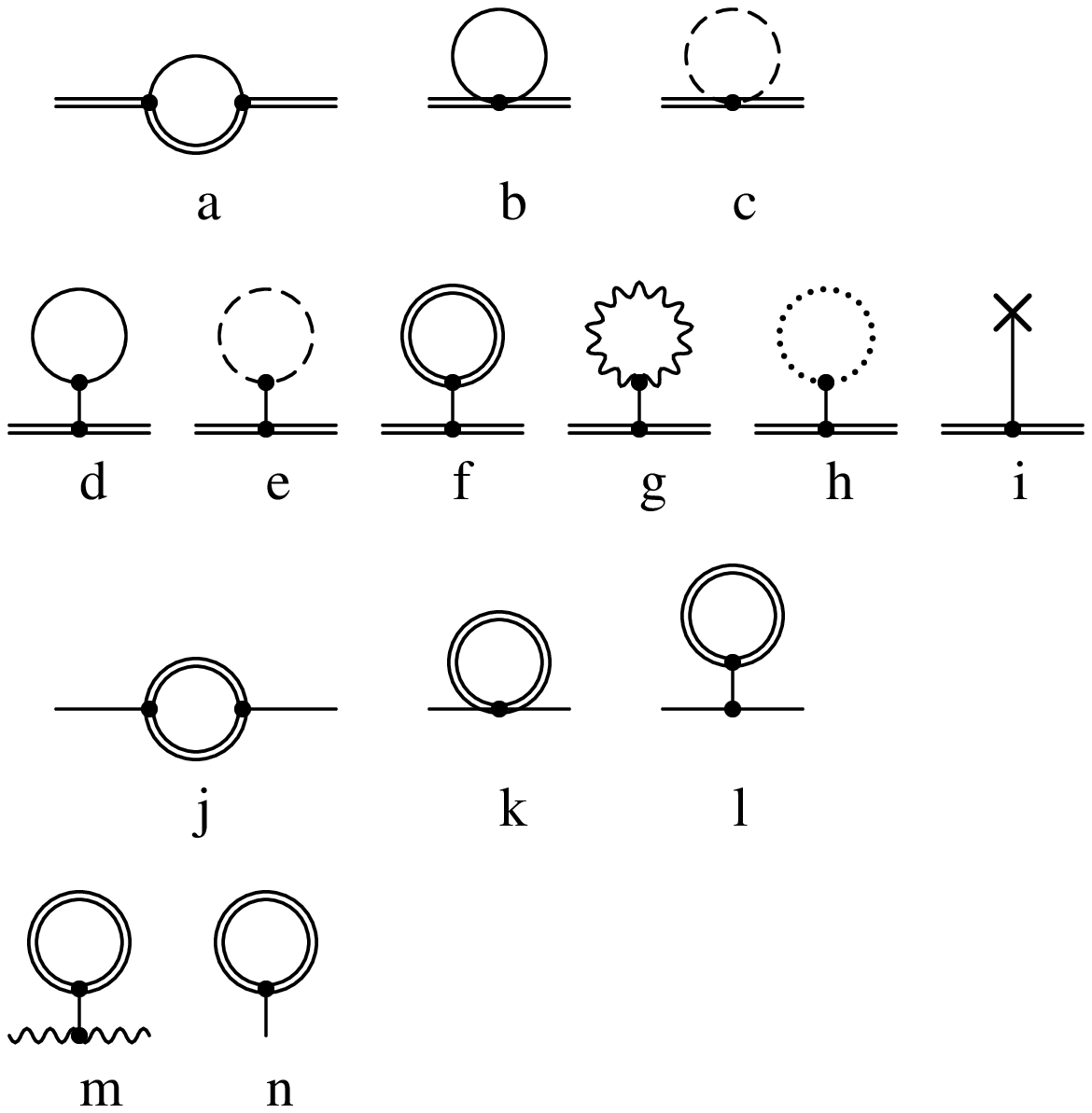}
}

\section*{Appendix B}

The gap equations for the masses in the gauge 
invariant resummation scheme read:
\bea
m^2&=&m_0^2+m g^2 f_B(m/M)+\frac{g^2}{2\pi}\left(-\frac{m_D}{2}+\frac{4m_D^2-m^2}{4m}
\textrm{arcth}\frac{m}{2m_D}\right),\\
M^2&=&M_0^2+g^2 M F_B(m/M)-\frac{3}{2\pi}({4mc\over g})^2\frac{1}{M}\textrm{arcth}
\frac{M}{2m_D}-\frac{3}{2\pi}cm_D,\\\nonumber
m_D^2&=&m_{D0}^2+\frac{g^2}{\pi}\left[-\frac{m_D}{2}-\frac{m}{2}+\left(m_D-
\frac{m^2}{4m_D}\right)\ln\frac{2m_D+m}{m}\right]-8v(\mu^2+\lambda
v^2)\frac{mc}{g M^2}+\\
&&\frac{1}{\pi}cM(1+6\frac{m^3}{M^3})+\frac{1}{\pi}({4mc\over g})^2
\left[\frac{3}{2}\frac{m_D}{M^2}-\frac{1}{2m_D}\ln\frac{2m_D+M}{M}\right],\\
&&
v(\mu^2+\lambda v^2)=-M^2\delta f_B(m/M)+\frac{3}{\pi}\frac{c}{g}m m_D.
\eea
where the $\delta f_B(z)=f_B(z)-\bar{f}_B(z)$ and $\bar{f}_B(z), f_B(z),
F_B(z)$ are defined by the eq. (24), (30) and (31). of \cite{bp94}:
\bea
\bar{f}_B(z) &=& {1\over \pi}\left[{63\over 64} \ln3 - {1\over 8} +
{1\over 32 z^3} - {1\over 32 z^2} + {1\over 8z} \right. \nn\\
&& \left.\quad + {3\over 4} z^2 - \left({1\over 64 z^4} - {1\over 16 z^2}
+ {1\over 8}\right)\ln(1+2z)\right]\ ,\label{barfw}\\
f_B(z) &=& {1\over \pi}\left[{63\over 64} \ln3 - {1\over 8} +
{1\over 32 z^3} - {1\over 32 z^2} - {1\over 16z}
- {3\sqrt\xi\over 16}\right. \nn\\
&& \left.\quad - \left({1\over 64 z^4} - {1\over 16 z^2}
+ {1\over 8}\right)\ln(1+2z)\right]\ ,\\
F_B(z) &=& {1\over \pi}\left[-\left({3\over 32} + {9\over 64}\ln3\right)
{1\over z^2} + {3\over 16}\left(1 - {3\over 2}\sqrt\xi\right){1\over z}
\right.\nn\\
&& \left.\quad - {3\over 8} z
- \left({3\over 8}z^2 - {3\over 16} + {3\over 64z^2}\right)
\ln{2z+1\over 2z-1}\right]\ .
\eea

\end{document}